%% file: Main_ISBI_new3.tex
\documentclass{article}
\usepackage{spconf,amsmath,graphicx}

\usepackage{enumitem}
\setlist{nosep, leftmargin=14pt}

\usepackage{mwe} 

\usepackage{spconf,amsmath,graphicx}

\newcommand{\ggg}[1]{\textcolor{gray}{#1}}
\newcommand{\ppp}[1]{\textcolor{blue}{#1}}
\newcommand{\yyy}[1]{\textcolor{orange!70}{#1}}
\usepackage{tabularray}
\UseTblrLibrary{booktabs}
\usepackage[colorlinks=true, allcolors=blue]{hyperref}
\usepackage[table]{xcolor} 
\usepackage{amsfonts}
\title{SynStitch: a Self-Supervised Learning Network for Ultrasound Image Stitching Using Synthetic Training Pairs and Indirect Supervision}
\name{\begin{tabular}{cc}
Xing Yao$^{1}$, Runxuan Yu$^{1}$, Dewei Hu$^{2}$, Hao Yang$^{1}$, Ange Lou$^{1}$, Jiacheng Wang$^{1}$, Daiwei Lu$^{1}$, Gabriel \\
Arenas$^{3}$, Baris Oguz$^{3}$, Alison Pouch$^{3}$, Nadav Schwartz$^{3}$, Brett C Byram$^{1}$, Ipek Oguz$^{1}$
\end{tabular}}
\address{$^{1}$ Vanderbilt University, $^{2}$ Mayo Clinic,  $^{3}$ University of Pennsylvania}

\begin{document}
%
\maketitle
\begin{abstract}
Ultrasound (US) image stitching can expand the field-of-view (FOV) by combining multiple US images from varied probe positions. However, registering US images with only partially overlapping anatomical contents is a challenging task. In this work, we introduce SynStitch, a self-supervised framework designed for 2DUS stitching. SynStitch consists of a synthetic stitching pair generation module (SSPGM) and an image stitching module (ISM). SSPGM utilizes a patch-conditioned ControlNet to generate realistic 2DUS stitching pairs with known affine matrix from a single input image. ISM then utilizes this synthetic paired data to learn 2DUS stitching in a supervised manner. Our framework was evaluated against multiple leading methods on a kidney ultrasound dataset, demonstrating superior 2DUS stitching performance through both qualitative and quantitative analyses. The code will be made public upon acceptance of the paper.

\end{abstract}
\section{Introduction}
\label{sec:intro}

Ultrasound (US) imaging is widely used in various diagnostic applications \cite{bano2024image,yao2024fnpc}. However, the limited field-of-view (FOV) in 2DUS poses challenges for visualizing anatomical structures and accurate downstream analysis \cite{fenster19963}. Image stitching methods allow generating images with larger FOVs by combining multiple US images from different probe positions. Existing stitching techniques \cite{10635432} can be generally categorized into feature-based methods, which use feature points and geometric descriptors, intensity-based methods, which rely on pixel intensity similarities, and tracker-based methods \cite{joy2024towards} that require additional hardware. Specifically in US stitching, Jyotirmoy et al.\ \cite{banerjee2015fast} developed a block-matching method for 3DUS. Gomez et al.\ \cite{gomez2019image} proposed a manifold-learning framework for 2DUS patch fusion and enhanced 3DUS stitching using a support vector machine to selectively utilize salient keypoints \cite{gomez2017fast}. Wright et al.\ \cite{wright2019complete} introduced an iterative spatial transformer network to align 3DUS volumes to a common atlas-space. They later developed a reinforcement learning strategy leveraging additional anatomical annotation for reward function construction \cite{wright2023fast}.
These methods are either not based on learning \cite{banerjee2015fast,gomez2017fast}, or they use an unsupervised learning scheme \cite{gomez2019image}, or require extensive manual annotation \cite{wright2019complete, wright2023fast} which prevents the generalizability to other US datasets. Therefore, the potential of deep learning (DL) in US stitching remains underexplored.





\begin{figure}[!t]
\centering
\includegraphics[width=0.4\textwidth]{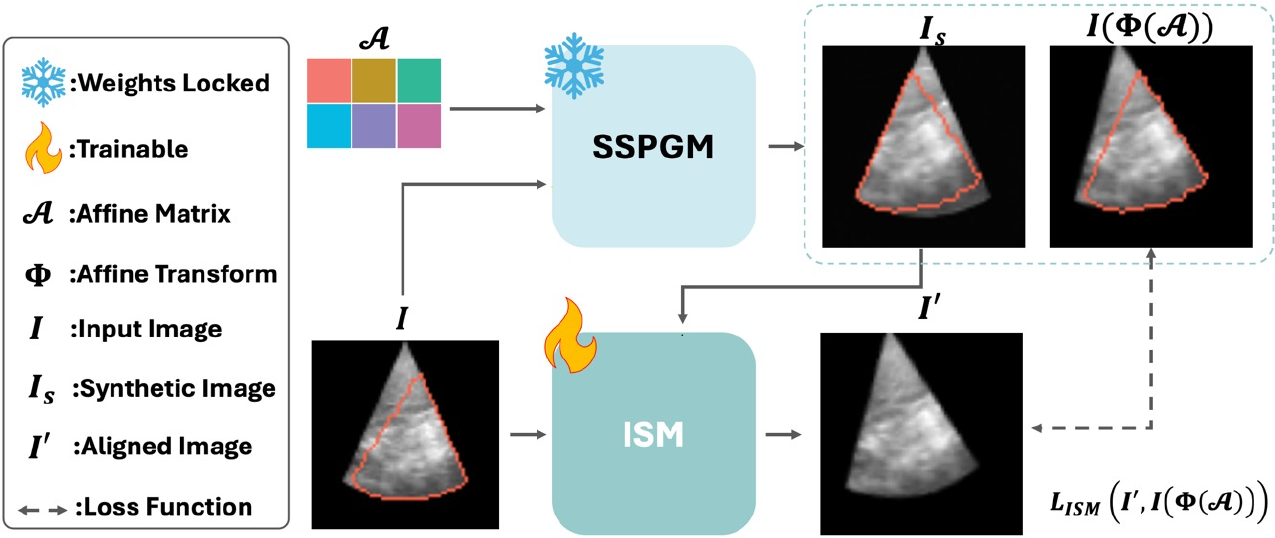}
\caption{\footnotesize SynStitch overview. We first train the SSPGM to generate a realistic 2DUS image \(I_s\) from an input image \(I\) with a random affine matrix \(\mathcal{A}\). Then we freeze the SSPGM and we train ISM on the synthetic stitching pairs.}
\label{whole_pipline}
\vspace{-5mm}
\end{figure}

\begin{figure*}
\centering
\includegraphics[width=1\textwidth]{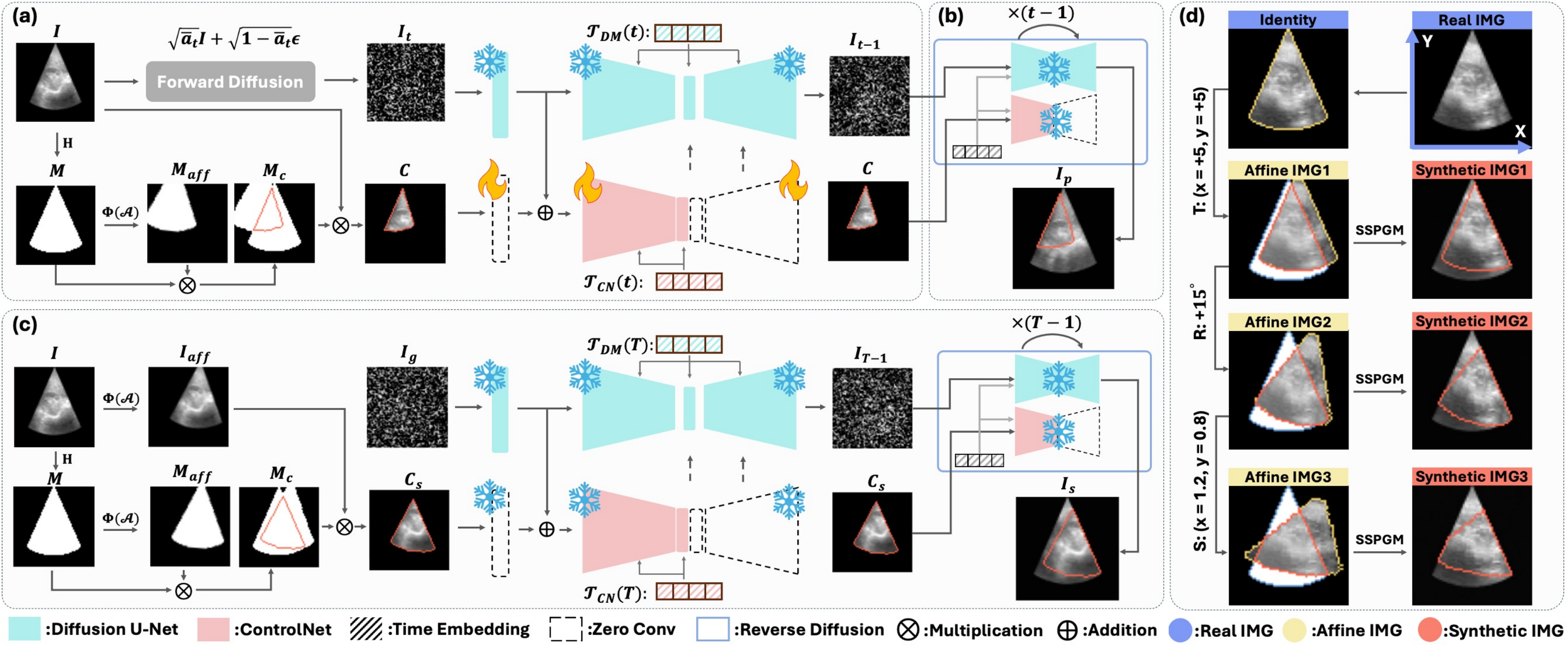}
\caption{\footnotesize Overview of the SSPGM. (a) Training: The SSPGM is trained to learn outpainting given the outpainting condition $C$. (b)  Inference: The trained SSPGM performs outpainting under the specified condition $C$. (c) Stitching pair generation: For a single input image $I$, a synthetic condition $C_{s}$ is generated and fed into the pre-trained SSPGM, which then generates  $I_{s}$ as a stitching pair for $I$, with associated affine matrix $\mathcal{A}$. (d) SSPGM results with custom affine matrices. \textbf{T}, \textbf{R}, \textbf{S} indicate translation, rotation, scaling. SSPGM can generate synthetic 2DUS images with a sequence of affine transformations. }


\label{SSPGM}
\vspace{-3mm}
\end{figure*}


In this work, we focus on stitching 2DUS images using affine transformations. Accurately stitching US images is challenging due to the uncontrolled motion of anatomical structures \cite{wright2023fast} and the presence of noise, shadows, and other artifacts. 
Importantly, the sector-shaped FOV poses significant challenges to both feature-based and intensity-based methods as it dominates loss/similarity functions.
Feature-based methods \cite{lowe1999object,lindenberger2023lightglue} can be misled by the sharp edges of the FOV. Similarly, intensity-based methods \cite{avants2008symmetric,hu2018label,balakrishnan2019voxelmorph,hoffmann2021synthmorph} tend to align the FOVs to minimize the intensity differences, thus neglecting the alignment of anatomical content. To mitigate the effects of the FOV, previous work proposed a patch-based method \cite{banerjee2015fast}. However, this approach faces challenges in achieving optimal patch selection especially in the narrow parts of the FOV.

To address these challenges, we propose SynStitch, a self-supervised learning (SSL) framework designed for 2DUS stitching (Fig.\ \ref{whole_pipline}). We begin by training a synthetic stitching pair generation module (SSPGM) based on the ControlNet architecture \cite{zhang2023adding,sharma2024controlpolypnet} that, given an input image \( I \), generates a 2DUS image \( I_s \) to form a realistic stitching pair \( (I, I_s) \). Ideal US stitching pairs should maintain identical FOV shape, visualizing overlapping regions but with partially dissimilar anatomical content due to the movement of the US probe. Therefore, SSPGM is trained to synthesize an $I_{s}$ that resembles $I$ in anatomical content but includes a random affine transformation $\mathcal{A}$ on the shared content (red contour in Fig.\ \ref{whole_pipline}). The  pair \((I, I_{s})\) is then fed into the image stitching module (ISM). ISM functions as a plug-and-play module, adaptable with any existing registration networks. During training, ISM aligns \(I\)  (moving image) to \(I_{s}\) (fixed image) in a supervised manner using true affine transform $\mathcal{A}$ by finding the matrix $\mathcal{A}'$ minimizing the difference between the aligned image \(I' = I(\Phi(\mathcal{A'}))\) and the ground truth \(I(\Phi(\mathcal{A}))\), where \( \Phi \) represents the affine coordinate transformation. ISM emphasizes the alignment of shared content, mitigating the impact of the FOV shape. We evaluate SynStitch against state-of-the-art feature-based and intensity-based methods on a 2DUS kidney dataset using both pixel-centric metrics and manually annotated keypoints. Our results show that SynStitch significantly ($p < 0.05$) enhances 2DUS stitching performance.
\section{Methods}
\label{sec:method}
\subsection{Synthetic Stitching Pair Generation
Module (SSPGM)}

\noindent \textbf{Overview:} The SSPGM (Fig.\ \ref{SSPGM}) is a patch-conditioned outpainting diffusion model that comprises a 2D Diffusion U-Net \cite{ho2020denoising} and a 2D ControlNet. ControlNet guides the pretrained Diffusion U-Net by embedding encoded conditions into the mid and decoder layers of U-Net.
The Diffusion U-Net is first trained unconditionally to learn to generate realistic 2DUS kidney images from random Gaussian noise input. After pretraining the Diffusion U-Net, the ControlNet is trained to outpaint an image \( I_p \) guided by a condition patch image \( C \) extracted from \( I \) (Fig.\ \ref{SSPGM}(a)-(b)). The outpainting result \( I_p \) features a sector-shaped FOV identical to the FOV of $I$ and is designed to replicate the same anatomical structure as \( I \) inside the condition patch $C$ (red contour). During the inference stage (Fig.\ \ref{SSPGM}(c)), a random affine transform $\mathcal{A}$ is generated to create an affine-transformed condition $C_s$. $C_s$ is used to generate a realistic synthetic image \( I_s \) that shares anatomical content with \( I \) but incorporates the affine transform $\mathcal{A}$. 
The implementation details are elaborated below.

\noindent \textbf{Diffusion U-Net Training:} The Diffusion U-Net training uses a forward diffusion process where a noisy image \( I_t \) is generated by incrementally adding Gaussian noise to  \( I \). This noisy image \( I_t \) at any given time step \( t \) is computed from \( I \) using the closed-form expression 
$I_t = \sqrt{\bar{\alpha}_t} I + \sqrt{1 - \bar{\alpha}_t} \epsilon$,
 where \(\bar{\alpha}_t = \prod_{s=1}^t \alpha_s\) is the cumulative product of noise scheduling coefficients \(\alpha_t\), and \(\epsilon \sim \mathcal{N}(0, I)\) is the Gaussian noise. In the subsequent denoising stage, the model \(\epsilon_\theta\), a denoising U-Net, takes the noisy image \( I_t \) as input to predict and subtract the noise \(\epsilon\) added to \( I \) at each timestep \( t \), aiming to minimize the loss defined as:



\begin{equation}
\mathcal{L_{DM}} = \mathbb{E}_{I, \epsilon \sim \mathcal{N}(0, I), t} \left[ \left\| \epsilon - \epsilon_\theta(I_t, t) \right\|^2 \right].
\label{eq:loss}
\end{equation}

\noindent \textbf{ControlNet Training:} The ControlNet is initialized by generating trainable copies of the encoder and middle layer weights from the Diffusion U-Net. As shown in Fig.\ \ref{SSPGM}(a), the generation of the training condition patch \( C \) starts with extracting a binary mask \( M = H(I) \) from the input image \( I \), using the thresholding function \( H \). A random affine matrix \( \mathcal{A} \) is then used to produce an affine-transformed binary mask \( M_{\text{aff}} = M(\Phi(\mathcal{A})) \). The overlapping region \( M_c = M \times M_{\text{aff}} \) is used as the condition mask (red contour). $M_c$ is used to create the outpainting condition \( C = M_c \times I \) for the ControlNet during training.
During the training phase, the noisy image \( I_t \) is input into the Diffusion U-Net, and the condition \( C \) is processed initially by the zero-convolution layers \cite{nichol2021improved}. The outputs are then combined with the original input and fed into the trainable copies of the ControlNet. Outputs from these trainable copies are then connected back to the Diffusion U-Net through zero-convolution layers. These layers are optimized gradually, improving training efficiency by minimizing noise introduction. 
Fig.\ \ref{SSPGM}(b) illustrates the inference process using the trained ControlNet with the condition patch \( C \), where \( I_p \) is expected to maintain the same anatomical structure as \( I \) inside the condition mask (red contour). Utilizing a forward diffusion process akin to that of Diffusion U-Net, the objective function for the patch-conditioned outpainting via ControlNet is:

\begin{equation}
\mathcal{L_{OP}} = \mathbb{E}_{I, C, \epsilon \sim \mathcal{N}(0, I), t} \left[ \left\| \epsilon - \epsilon_\theta(I_t, C, t) \right\|^2 \right],
\label{eq:cnloss}
\end{equation}


\noindent \textbf{Inference Stage with Condition Generation:} As illustrated in Fig.\ \ref{SSPGM}(c), the inference stage employs a distinct patch condition generation strategy to synthesize images with relative motion to \( I \). Initially, a random affine matrix \( \mathcal{A} \) is applied to transform both \( I \) and its mask \( M \), yielding \( I_{\text{aff}} = I(\Phi(\mathcal{A})) \) and \( M_{\text{aff}} = M(\Phi(\mathcal{A})) \). The overlap region between \( M \) and \( M_{\text{aff}} \) is then extracted to create the condition mask \( M_c \), which is used to derive the condition image \( C_s = I_{\text{aff}} \times M_c \) from the transformed input \( I_{\text{aff}} \). Unlike the training  condition \( C = I \times M_c \), the inference condition \( C_s =  I_\text{aff} \times M_c\)  is derived from the affine transformed image \( I_\text{aff} \). To synthesize the image \( I_s \) from \( C_s \), we begin by sampling Gaussian noise \( I_g \) as the input to the pretrained ControlNet. This is followed by \( T \) steps of denoising, guided by \( C_s \), culminating in the synthetic image \( I_s \). This image \( I_s \) not only retains the overlapping anatomical structures in \( I \) but also incorporates the affine transform. Thus, it forms a coherent stitching pair \( (I, I_s) \) with the input image, underpinned by the known ground truth affine matrix \( \mathcal{A} \) that aligns \( I \) with \( I_s \).



\subsection{Image Stitching Module (ISM)}

The Image Stitching Module (ISM) (Fig.\ \ref{whole_pipline}) is a trainable neural network designed to learn 2DUS image stitching from a synthetic stitching dataset \( \mathcal{S} = \{ (I^k, I^k_s, \mathcal{A}^k) \}_{k=1}^K \) in a supervised manner. Given the input source (\( I^k \)) and target (\( I^k_s \)) images, we optimize the trainable parameters $\theta$ of  the ISM \( \mathcal{M} \)  to minimize the objective function 
$\mathcal{L_{ISM}} = \mathbb{E}_{(I^k, I^k_s, \mathcal{A}^k) \in \mathcal{S}} \left[ \left\| \mathcal{M}_{\theta}(I^k, I^k_s) - I^k(\Phi(\mathcal{A}^k)) \right\|^2 \right].$




\section{Experimental Design}

\begin{figure*}[t]
\centering
\includegraphics[width=1\textwidth]{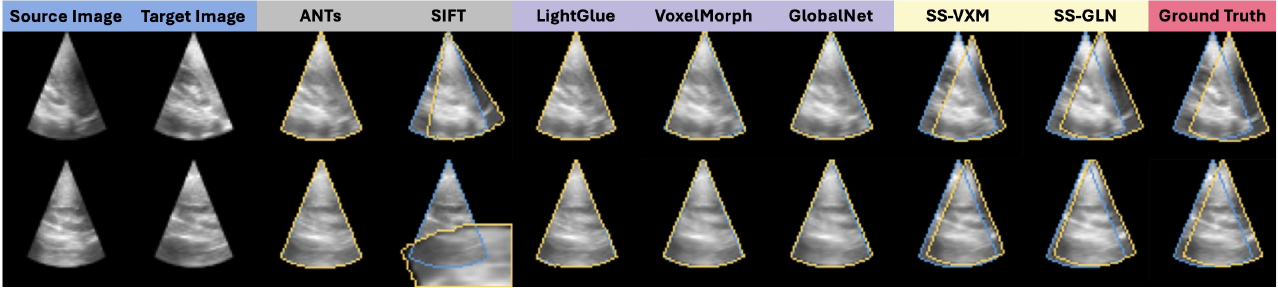}
\caption{\footnotesize Stitching results from randomly selected samples from two RealStitch subjects. Yellow contours:  registered source images;  blue contours:  fixed images. Blue tags: source and target images; gray tags: conventional methods; purple tags: DL methods;  yellow tags:  the two variants of our proposed model, SynStitch-VXM (SS-VXM) and SynStitch-GLN (SS-GLN); red tag: ground truth. ANTs and DL methods only align the FOV and fail to align anatomy. SIFT attempts to align anatomy but is not accurate. Our method produces robust results.}
\label{stitch_results}
\vspace{-3mm}
\end{figure*}

\input{resultsTable}

\noindent \textbf{\underline{Dataset:}} 
We acquired free-hand 4D ultrasound (4DUS) videos of kidneys from 14 healthy subjects. The probe was maneuvered to capture 3D frames of the kidneys from various angles and positions. Each subject contributed 3-4 4D videos, with each video comprising 124 to 125 3D frames of dimensions \(128 \times 128 \times 128\) and a spacing of \(1 \times 1 \times 1\). Subjects were randomly divided for training, validation, and testing in a 9:3:5 ratio. This data was leveraged in 5  datasets as described below.

\noindent \underline{\textit{(a) SSPGM Training Dataset:}} To train the SSPGM, we extracted two central 2D slices with a sector-shaped FOV from the coronal and axial planes of each 3D volume, resulting in 7478 training  and 2494 validation images. All extracted 2D images were resized to \(64 \times 64\) and normalized to  [0, 1].

\noindent \underline{\textit{(b) RealCurated Dataset:}} 
Images from the SSPGM Training dataset were curated such that either only coronal or only axial slices are selected from a given 3D volume, to ensure the probe motion is noticeable in the images. This selection produced 3739 training images and 1247 validation images.

\noindent \underline{\textit{(c) SynStitch Dataset:}} 
For training and validating ISM, images from the RealCurated Dataset were used to generate corresponding synthetic stitching pairs and ground truth affine matrices using the pretrained SSPGM. 

\noindent \underline{\textit{(d) RealTrain Dataset:}} To train and validate learning-based affine registration baselines, we dynamically extract 3739 real stitching pairs from the RealCurated Dataset  for each epoch. Each pair consists of two images randomly selected from the same subject, with a frame index difference of less than 20 to ensure noticeable relative motion between them.

\noindent \underline{\textit{(e) RealStitch Dataset:}} 60 real 2DUS stitching pairs were derived from 5 subjects, with each pair featuring 10 manually annotated keypoints and manually aligned results.



\noindent \textbf{\underline{{Implementation Details:}}} We implemented the 2D Diffusion U-Net and ControlNet components of the SSPGM using MONAI \cite{pinaya2307generative}, built on a 3-level U-Net architecture with encoder channels at 128, 256, and 256. Both components are trained 200 epochs on an A6000 GPU, with a learning rate of \(1e^{-5}\) and a batch size of 64, using DDPM \cite{ho2020denoising} with 1000 sampling steps. For ControlNet training, augmentation included random translations (\([-24, 24]\)), rotations (\([- \frac{\pi}{12}, \frac{\pi}{12}]\)), and scaling (\([0.9, 1.1]\)). The \textbf{SynStitch Dataset} was generated with smaller translations (\([-8, 8]\)) and rotations (\([- \frac{\pi}{24}, \frac{\pi}{24}]\)) to mimic subtle probe movements during  acquisition.

\noindent \textbf{\underline{{Baseline methods and evaluation metrics:}}} 
We evaluate the  performance of our SynStitch method on the \textbf{RealStitch Dataset} against a variety of approaches. These include two conventional state-of-the-art methods, ANTs-Affine \cite{avants2008symmetric} and 2D SIFT \cite{lowe1999object} with RANSAC \cite{fischler1981random}; the pretrained feature-matching method LightGlue \cite{lindenberger2023lightglue}, which utilizes SuperPoint \cite{detone2018superpoint} as the keypoint extractor; and two SSL-based affine registration methods, GlobalNet \cite{hu2018label} and VoxelMorph-Affine \cite{balakrishnan2019voxelmorph,hoffmann2021synthmorph}, trained on the \textbf{RealTrain Dataset} with a learning rate of \(1e^{-5}\) and a batch size of 128 for 500 epoches.  GlobalNet (GLB) and VoxelMorph (VXM) were integrated as the ISM within the SynStitch framework, resulting in the variants SynStitch-GLB and SynStitch-VXM, which are both trained on the \textbf{SynStitch Dataset} with the same hyper-parameter as their counterparts. We report pixel-centric metrics (MSE, SSIM, NCC) and the RMSE of manually annotated keypoints. 

\section{Results}


\noindent \textbf{\underline{{Synthetic Image Generation Results:}}} Fig.\ \ref{SSPGM}(d) demonstrates the SSPGM's capability to generate realistic 2DUS images using arbitrary affine matrices. Starting with a real input image, we iteratively apply translation (\(T\)), rotation (\(R\)), and scaling (\(S\)). The condition patches (red contours)  guide the SSPGM in synthesizing realistic images. The resulting images retain the structures within the condition patches and seamlessly extend realistic structures beyond these contours, effectively handling a considerable range of motion.

\noindent \textbf{\underline{{Real 2DUS Stitching Results:}}} Fig.\ \ \ref{stitch_results} shows  stitching results for random samples from two subjects from the \textbf{RealStitch Dataset}. The results reveal that ANTs and all learning-based baseline methods typically align the sector-shaped FOVs of the input images but fail to accurately stitch the anatomical structures. While SIFT outperforms other baseline methods in the first row, it still exhibits inaccurate rotation as it is misled by high-intensity regions at the top of the FOV. In the second row, SIFT is severely compromised by  motion and shadows, leading to mismatched features. In contrast, the proposed SynStitch framework accurately aligns the source to the target image, handling both large (first row) and smaller motions (second row) effectively. Although SynStitch-VXM shows comparable performance to SynStitch-GLN, it exhibits slight shearing relative to the ground truth.

Table \ref{table_test1} presents quantitative results on the RealStitch Dataset. Consistent with qualitative results, all the baseline methods except SIFT share a similar performance that is dominated by aligning the FOV rather than anatomical content. While SIFT achieves a better RMSE compared to these other baselines, it exhibits the worst MSE and highest standard deviation across all metrics, indicating its unstable performance on this task, which aligns with observations from Fig.\ \ref{stitch_results}. SynStitch-VXM and SynStitch-GLN significantly  outperform others in all evaluation metrics. This shows SynStitch is highly effective in optimizing existing approaches such as VoxelMorph and GlobalNet for  2DUS stitching. While SynStitch-VXM has slightly lower performance than SynStitch-GLN, both exceed all baseline methods.

\section{Conclusion}
In this study, we introduced SynStitch, a self-supervised framework for 2DUS stitching. SynStitch generates realistic 2DUS stitching pairs with ground truth affine transform from a single input image, and learns stitching from this synthetic dataset. Our evaluations against state-of-the-art feature-based and intensity-based methods on a 2DUS kidney dataset showed significant improvements in both qualitative and quantitative performance. These results further illustrate that SynStitch can integrate with existing SSL registration frameworks, enhancing their effectiveness in 2DUS stitching tasks. Future work will focus on validating SynStitch across additional models and datasets, and  other medical imaging domains such as fetoscopy \cite{honeywell2024real} and retinal imaging \cite{wang2024retinal}. 

\noindent \textbf{Acknowledgments.} This work is supported, in part, by NIH R01-HD109739, R01-HL156034, T32-EB021937, and the Vanderbilt Advanced Computing Center for Research and Education.  
\noindent\textbf{Compliance with Ethical Standards.}
The study was approved by the Vanderbilt  IRB (IRB No.\ 222195).


\bibliographystyle{IEEEbib}
\bibliography{strings,refs}

\end{document}

%% file: resultsTable.tex
\begin{table*}[t]
\centering
\caption{Quantitative results from RealStitch. \textbf{Bold}: best, \underline{underline}: second-best, *significant improvements (paires t-test, p\textless 0.05), \ggg{gray}: conventional methods,\ppp{purple}:DL methods,\yyy{yellow}:proposed methods. MSE: mean square error; SSIM: structural similarity index measure; NCC: normalized cross-correlation; RMSE: root mean square error of keypoints.}
\label{table_test1}
\begin{small} 
\begin{tblr}{
  colspec={|l|c|c|c|c|c|c|c|},
  rows={m},
  hline{1,2,3,4,5}={solid},
  vline{2-6}={solid},
  colsep=1.7pt, 
  rowsep=0.6pt, 
  column{1}={bg=white!10},
  column{2}={bg=gray!10},
  column{3}={bg=gray!10},
  column{4}={bg=blue!10},
  column{5}={bg=blue!10},
  column{6}={bg=blue!10},
  column{7}={bg=yellow!10},
  column{8}={bg=yellow!10},
}
\toprule
Metrics & ANTs\cite{avants2008symmetric} & SIFT\cite{lowe1999object} & LightGlue\cite{lindenberger2023lightglue}  & VoxelMorph\cite{balakrishnan2019voxelmorph}  & GlobalNet\cite{hu2018label}  & SynStitch-VXM  & SynStitch-GLN \\
\midrule
MSE(x100)\(\downarrow\) & 2.89 \(\pm\) 1.64 & 3.03 \(\pm\) 2.75 & 2.82 \(\pm\) 1.68 & 2.73 \(\pm\) 1.60 & 2.81 \(\pm\) 1.63 & \underline{1.90 \(\pm\) 1.38}* & \textbf{1.54 \(\pm\) 1.14}* \\
SSIM\(\uparrow\) & 0.61 \(\pm\) 0.09 & 0.62 \(\pm\) 0.12 & 0.62 \(\pm\) 0.10 & 0.63 \(\pm\) 0.10 & 0.62 \(\pm\) 0.10 & \underline{0.69 \(\pm\) 0.12}* & \textbf{0.71 \(\pm\) 0.11}*\\
NCC\(\uparrow\) & 0.67 \(\pm\) 0.17 & 0.67 \(\pm\) 0.21 & 0.68 \(\pm\) 0.18 & 0.69 \(\pm\) 0.17 & 0.68 \(\pm\) 0.17 & \underline{0.78 \(\pm\) 0.15}* & \textbf{0.82 \(\pm\) 0.13}*\\
RMSE\(\downarrow\) & 5.13 \(\pm\) 2.63 & 4.93 \(\pm\) 3.35 & 5.09 \(\pm\) 2.67 & 5.01 \(\pm\) 2.62 & 5.21 \(\pm\) 2.62 & \underline{3.75 \(\pm\) 2.20}*  & \textbf{3.36 \(\pm\) 1.94}*\\
\bottomrule
\end{tblr}
\end{small}
\vspace{-3mm}
\end{table*}
